%% file: resim.tex
\documentclass[usenatbib]{mn2e}
\usepackage{psfig}

\usepackage{latexsym}

\newcommand{\keV}{\mbox{keV}}
\newcommand{\Lx}{\mbox{$L_{\rm X}$}}

\newcommand{\Msun}{\mbox{${\rm M}_{\odot}$}}
\newcommand{\Tx}{\mbox{$T_{\rm X}$}}

\title[The merger history of clusters] {The merger history of clusters
and its effect on the X-ray properties of the intracluster medium}

\author[David R. Rowley et al.]
{David R. Rowley\thanks{E-mail: d.r.rowley@susx.ac.uk}, 
Peter A. Thomas and Scott T. Kay\\
Astronomy Centre, Department of Physics and Astronomy, University of
Sussex, Brighton, BN1 9QH}

\date{\today}

\pagerange{\pageref{firstpage}--\pageref{lastpage}}
\pubyear{2003}

\begin{document}
\journal{Preprint astro-ph/0310493} 

\maketitle

\label{firstpage}

\begin{abstract}

We investigate the growth over time of 20 massive ($>$ 3 \keV)
clusters in a hydrodynamical simulation of the $\Lambda$CDM cosmology
with radiative cooling.  The clusters show a variety of formation
histories: some accrete most of their mass in major mergers; others
more gradually.  During major mergers the long-term
(temporally-smoothed) luminosity increases such that the cluster moves
approximately along the \Lx-\Tx\ relation; between times it slowly
decreases, tracking the drift of the \Lx-\Tx\ relation.  We identify
several different kinds of short-term luminosity and temperature
fluctuations associated with major mergers including double-peaked
mergers in which the global intracluster medium merges first (\Lx\ and
\Tx\ increase together) and then the cluster cores merge (\Lx\
increases and \Tx\ decreases).  At both luminosity peaks, clusters
tend to appear spherical and relaxed, which may lead to biases in
high-redshift, flux-limited samples.  There is no simple relationship
between scatter in the \Lx-\Tx\ relation and either recent or overall
merger activity or cluster formation time.  The scatter in the \Lx-$M$
and \Tx-$M$ relations is reduced if properties are measured within
$R_{500}$ rather than $R_{\rm{vir}}$.
\end{abstract}

\begin{keywords}
galaxies: clusters: general
\end{keywords}

\section{Introduction}
Clusters of galaxies are the largest virialized structures in the
Universe and mergers between them are among the most energetic
events. This paper presents results from resimulations of 20 massive
clusters to investigate the effects of the merger history on the X-ray
properties of the intracluster medium (ICM).

In recent years high-quality observations of the ICM of distant
clusters of galaxies have become available.  This has enabled
cosmological evolution studies to be carried out on clusters by
looking at, for example, the high-redshift \Lx-\Tx\ relation
\citep{FJS00,BRT01,HSS02,VVM02,NSH02}.  This paper will look in detail
at the merger history of clusters of galaxies and how this affects
their X-ray properties.  Previous observational studies of merging
clusters of galaxies include \citet{MGD02}, \citet{MJE03} and
\cite{RSK03}, and \citet{ASF02} found evidence that the hottest
cluster known (RX J1347.5-1145) is undergoing a merger.  We find
examples in our simulated clusters that mimic these observed clusters.

The merger rates of simulated Cold Dark Matter (CDM) haloes were first
investigated by \cite{LaC94} and \cite{NFW95a}.  The results were
found to be in agreement with the analytical model of \cite{LaC93}.
\citet{WBP02} and \citet{ZMJ03} looked at the effects of mergers and
accretion on the structure of individual CDM haloes and found that halo
concentration tends to increase with increasing formation redshift.

To investigate the effects of mergers on the ICM, \citet*{PTC94}
simulated the collision of pairs of relaxed cluster haloes and
compared these with earlier dark-matter only simulations
\citep*{PTC93}.  They found that the entropy structure of both the
dark matter and the gas is relatively unchanged during the merger,
although there is a small tendency to transfer energy from the former
to the latter in the core of the system.  The time evolution of the
X-ray properties of merging clusters were investigated by
\citet{RiT02} in higher resolution simulations that also included
radiative cooling.  They showed that the X-ray temperature and
luminosity can temporarily increase by a large factor during a merger;
the central entropy increase after the merger was found to be a strong
function of subcluster mass and impact parameter.

The simulations described above started with isolated, relaxed
clusters and investigated different impact parameters and mass
ratios. However, real clusters in a cosmological environment are more
complicated.  They will have some amount of substructure, may be
rotating and could collide with more than one subclump at a time.
Therefore to find out what kind of mergers clusters tend to undergo
and the effect this has on the complex ICM, full cosmological
simulations need to be carried out.
A first step in this direction was undertaken by \citet*{ENF98}.  They
undertook resimulations of individual clusters extracted from a
dark-matter simulation and showed that the evolution of bulk
properties of clusters varied from cluster to cluster.  The
simulations that we describe in this paper are similar in spirit but
have more particles, include radiative cooling of the gas component,
and have a much higher time resolution with which to follow the
development of the X-ray properties.
 
We should also mention a couple of other recent studies.
\citet{RSR02} looked at the observational bias that can be induced by
the temporary enhancements in luminosity and temperature in merging
clusters on determinations of $\sigma_8$ and the number density of
high-redshift clusters.  \citet{MBL03} undertook radiative simulations
of a sample of two clusters and showed that the traditional model of
smooth accretion onto clusters is inaccurate in that clusters accrete
gas in subclumps which bring precooled gas direct to the core of a
cluster.  Our work supports both of these ideas.

The rest of this paper is organized as follows.  In Section~2 the
properties of the simulations and the resimulation technique will be
described.  Section~3 will investigate the mass, luminosity and
temperature evolution of the resimulated clusters, with particular
types of event being identified and examples of these looked at in
detail.  The link between scatter in the scaling relations and merger
history or substructure will be investigated in Section~4.  Finally,
in Section~5, we summarise our results and compare the features seen
in the simulations to observations.

\section{The simulations}

In a previous paper, \citet[hereafter MTKP02]{MTK02}, X-ray scaling
relations were presented for simulated catalogues containing over 400
groups and clusters.  However, because of the volume simulated, only 7
of these had virial temperatures in excess of 3\,keV.  This study
simulates a further 20 such clusters by resimulating regions within a
larger box, with the high-resolution regions having simulation
parameters identical to that of the
earlier run.

Three simulations were undertaken with parameters as shown in
Table~\ref{tab:simpar}.
\begin{table}
\caption{Simulation parameters for the three runs: name; effective
number of particles; particle mass in units of $h^{-1}$\Msun;
softening in units of $h^{-1}$kpc and particle species simulated.}
\label{tab:simpar}
\begin{center}
\begin{tabular}{lcccl}
Name& Particles& Mass& Soft.& Particle types\\\hline
LDM&  160$^3$& $1.9\times10^{11}$& 50& Dark matter \\
HDM&  320$^3$& $2.4\times10^{10}$& 25& Dark matter \\
HGAS& $2\times320^3$& $2.4\times10^{10}$& 25& Dark matter + gas
\end{tabular}
\end{center}
\end{table}
The effective number of particles is the number of particles needed to
fill the entire simulation volume at the highest resolution in the
box.  In the simulation HGAS there were equal numbers of gas and dark
matter particles in the high-resolution regions. The mass resolution
for HGAS refers to the sum of the two species.

\subsection{The resimulation technique}

We first generated a low-resolution run (LDM) with 160$^3$ dark-matter
particles.  This was run to completion (from $z=49$ to $z=0$) and the
20 largest clusters identified using the technique described in
MTKP02.  All the particles within 2 virial radii of the cluster
centres were identified.  The locations of all these particles on the
initial comoving grid (i.e.~before the Zel'dovich approximation is
applied) were noted and, along with the neighbouring grid points, were
flagged as requiring high resolution.  Just under 4 per cent of the
box was included in the high-resolution mask.

A second set of initial conditions was then generated on a 320$^3$
grid.  The original waves used in the LDM simulation were used again
with additional new waves generated for frequencies between the LDM \&
HDM Nyquist frequencies.  The high- and low-resolution initial
conditions were then combined with the former being used within the
previously-identified mask and the latter elsewhere.  This resulted in
5\,191\,535 dark-matter particles (as compared to 4\,096\,000 in the
original low-resolution simulation).

Finally a third set of initial conditions was generated that included
both gas and dark matter in the high-resolution regions.  To do this,
a gas particle was added at the location of each high-resolution
dark-matter particle.  The masses of the two were adjusted to give the
same total mass as before and to give the universal dark matter to
baryonic matter mass ratio.  The dark-matter and gas particles were
given identical displacements and velocities.  The total number of
particles for this run was 6\,443\,575.

The hydrodynamics simulation used parameters identical to that for the
{\it Radiative} model of MTKP02.  In particular, it included radiative
cooling with a time-varying metalicity $Z(t) = 0.3\,(t/t_0)\,Z_\odot$,
where $t_0$ is the current age of the universe (approximately 12.8~Gyr
for this cosmology), using cooling tables from \citet{SuD93}. Neither
preheating nor feedback were included in this run.

The simulations were run on 64 processors on the Cray T3E at the EPCC,
using a parallel form of the AP$^3$M SPH code {\small{HYDRA}}
\citep{CPT95}\footnote{\small{HYDRA} is available from {\tt
http://hydra.susx.ac.uk/}}.  The box was $200\,h^{-1}$\,Mpc across and
used cosmological parameters: $h_0 = 0.71$; $\sigma_8 =0.9$;
$\Omega_{\rm{m}} =0.35$; $\Omega_{\rm{v}} =0.65$ and
$\Omega_{\rm{b}}=0.038$.  The shape parameter for the fluctuation
spectrum was fixed at $\Gamma=0.21$ using the fitting function of
\citet{BoE84}.  The gravitational softening was set to
100/(1+z)$\,h^{-1}$kpc until $z=3$, after which it was held fixed at
25$\,h^{-1}$kpc. The initial redshift was $z=49$ and the
highest-resolution run required 3706 timesteps to evolve to $z=0$.

\subsection{Cluster properties}

The properties of the resimulated clusters at $z=0$ are listed in
Table~\ref{tab:clus}.  \input clus.tab The virial radius is defined as
the radius of a sphere, centred on the densest dark-matter particle,
that encloses a mean density of 317 (specifically
$178\,\Omega_{\rm{m}}^{-0.55}$, \citealt{ENF98}) times the background
density (111 times the critical density).  The virial mass, $M_{\rm
vir}$, is the mass enclosed by this sphere and the virial temperature
is the mean specific energy (kinetic plus thermal) of the dark matter
and gas, multiplied by $\mu m_H/k$, where $\mu m_H=10^{-24}$g is the
mean molecular mass and $k$ is the Boltzmann constant.  The X-ray
temperature is weighted by emission in the soft X-ray band
(0.3--1.5\,keV), excluding emission from within $50\,h^{-1}$kpc
(i.e.~physical, not comoving) of the cluster centre (hereafter
referred to as cooling-flow corrected emission).  The soft-band X-ray
luminosity is converted into an estimated bolometric luminosity using
the procedure described in MTKP02.  The substructure statistic is
defined as the separation between the centroid of the mass and the
location of the densest dark-matter particle, in units of the virial
radius \citep{TCC98}.  The formation expansion factor, $a_{\rm f}$, is
a characteristic formation epoch for the cluster as defined in
Section~3.1.

Note that, after resimulation, the virial mass and temperature of the
final cluster in the list dropped significantly.  This is because the
cluster is in the process of merging with a subcluster that lay just
inside the virial radius in the original low-resolution run but has
moved just outside it in the high-resolution run.  Even so, this
cluster still has over 13\,000 particles each of gas and dark matter
within the virial radius at the final time. The largest cluster has
approximately 73\,000 particles of each type at the end.

\subsection{Testing}

The mass of dark-matter particles in the low-resolution regions
exceeds that of those in the high-resolution regions by a factor of $\sim$9
(and of the gas particles by a factor of $\sim$74). It is therefore
necessary to ensure that the clusters are not affected by these high
mass particles which could cause unphysical two-body relaxation (since
the softening is the same for all the particles). The clusters were
examined to discern if there were any low-resolution particles within
their virial radii at any output time.  It was found that only one of
the clusters presented was at all affected.  This was Cluster~13 which
temporarily has up to 2 low-resolution dark-matter particles within
the virial radius during a late-time merger.  We do not expect this to
affect the results and we choose to ignore it.

\begin{figure}
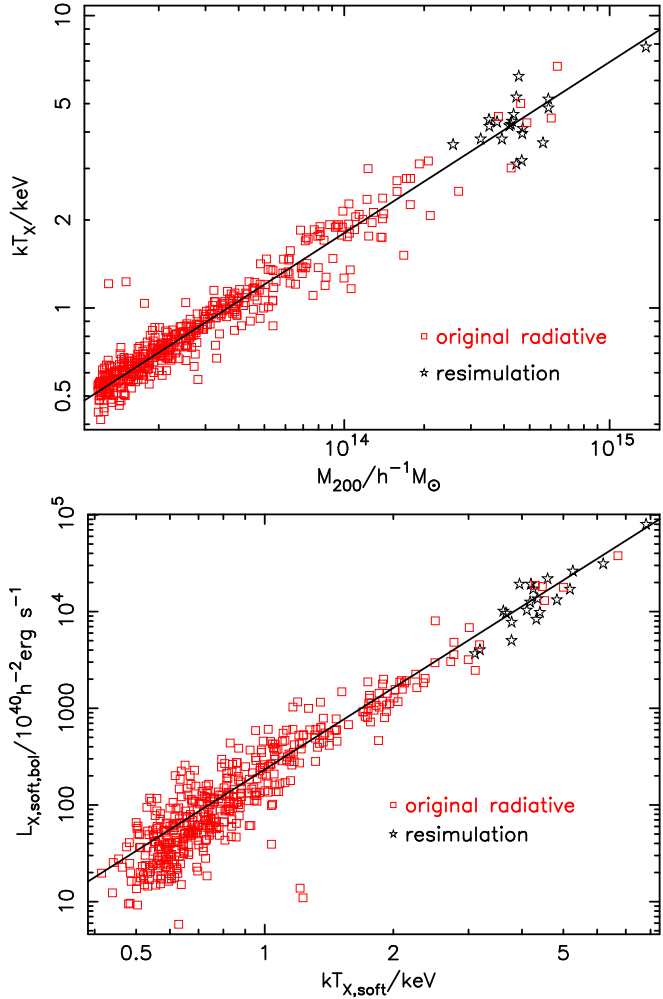

\centerline{\psfig{file=txsoftm_rfixed.ps,width=8.7cm,angle=270}}
\centerline{\psfig{file=lxboltxsoft_rfixed.ps,width=8.7cm,angle=270}}
\caption{Cooling-flow corrected $\Tx-M_{200}$ (upper panel) and
  $\Lx-\Tx$ (lower panel) relation in the soft X-ray band (luminosity
  converted into an estimation of the bolometric luminosity) for the
  clusters from MTKP02 (squares) and those from simulation HGAS
  (stars).  The lines have a slope of 0.59 (upper panel) and 2.80
  (lower panel).}
\label{fig:txsoftm_rfixed}
\end{figure}

As a test of our method the clusters were compared to those from the
{\it Radiative} simulation of MTKP02.  A comparison between the
temperature-mass relation and the luminosity-temperature relation for
the two is shown in Fig.~\ref{fig:txsoftm_rfixed}
To be consistent with MTKP02, the extent of the clusters is defined by
a sphere that encloses a mean density of 200 times the critical
density.  It is clear that the new clusters are consistent with the
previous relations but extend them to higher mass and temperature.

\begin{figure}
\centerline{\psfig{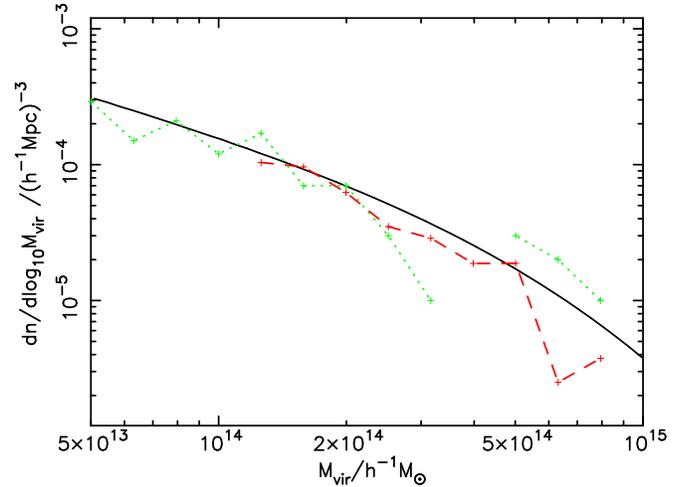}}
\caption{The mass function of clusters in the LDM simulation (dashed
  line) and in the simulations of MTKP02 (dotted line) compared to the
  prediction from \citet{JFW01} (solid line).  The most-massive
  cluster lies off the right-hand-side of the plot and has not been
  included in the figure as there are numerous empty bins between it
  and the second most-massive cluster.}
\label{fig:massfn}
\end{figure}

One surprising result, however, is that the number of clusters above a
virial temperature of 3\,keV in the new runs is far fewer than 8 times
the number in the original MTKP02 simulation (20 as compared to 7)
which was one-eighth of the volume.  To test our normalization, we
compare in Fig.~\ref{fig:massfn} the mass-function of the simulated
clusters with that predicted by \citet{JFW01} from a compilation of a
large number of $N$-body simulations.  The 20 high-resolution clusters
are contained in the final three bins of the plot (except for the most
massive cluster that lies off the right-hand-side of the plot).  In
the same mass-range, 3 clusters are expected in the MTKP02 simulation
but 6 are found.  We put this down partly to chance and partly due to
the fact that simulations do not sample waves correctly on scales
comparable to the box-size.

\section{The time evolution of Mass, X-ray luminosity and temperature}

In this section we look at the change in the mass, X-ray luminosity
and X-ray temperature of the clusters as they evolve.  For simplicity
and for ease of physical interpretation, we use the bolometric
(i.e.~not the soft-band used above) emission from within the virial
radius in the rest-frame of the cluster.

The time resolution is defined by the light crossing time of half the
box, that is $0.5\times200\,h^{-1}{\rm{Mpc}}\times a/c \approx 4.6
\times 10^{8}\times a$\,yr, where $c$ is the speed of light and
$a=1/(1+z)$ is the expansion factor.  These times were chosen to match
previous simulations which were used to obtain light-cones.  This
gives sufficient resolution to crudely resolve the mergers for the
clusters as a whole, although finer time-resolution would have allowed
us to examine accretion into the cores of the clusters in more detail.

\subsection{The growth of mass}

\begin{figure*}
\centerline{\psfig{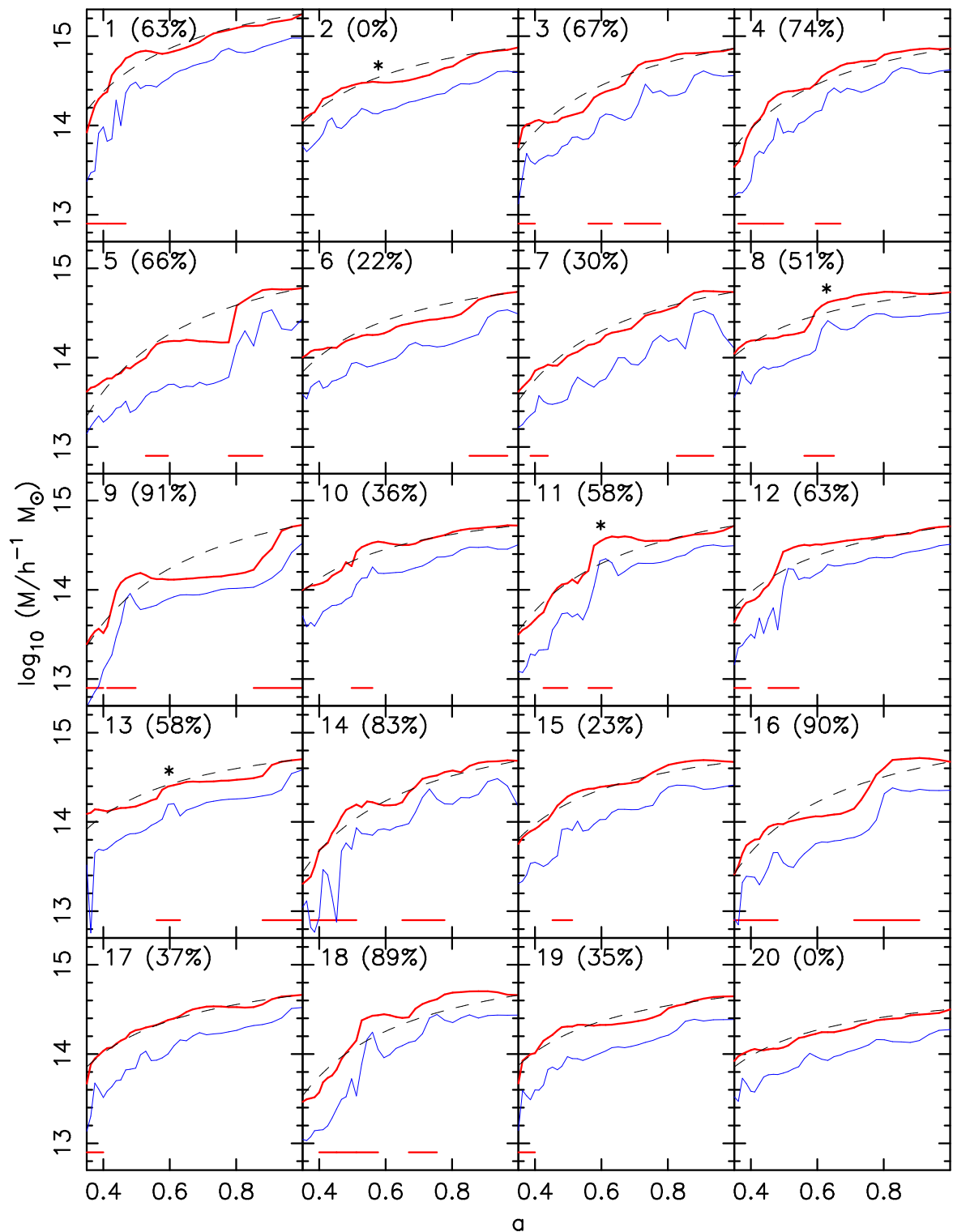}}
\caption{$M_{\rm{vir}}$ (thick line) and $M_{500}$ (thin line) versus
  expansion factor for the clusters.  The dashed curves are a
  one-parameter fit to the functional form
  $M(a)=M_0\,e^{2a_{\rm{f}}(1-\frac{1}{a})}$ introduced by
  \citet{WBP02}, where $a_{\rm{f}}$ is listed in Table~2. The
  asterisks indicate events which will be investigated in detail
  below. The lines along the bottom of each box indicate the times
  during which the cluster is deemed to be undergoing a major merger
  (as defined in the text).  The numbers in parentheses indicate the
  fractional logarithmic mass accreted during major mergers.}
\label{fig:mevol}
\end{figure*}

Fig.~\ref{fig:mevol} shows the time-development of the mass enclosed
within $R_{\rm{vir}}$.  The most-massive cluster is located in the
top-left panel of this figure (and also in Figs~\ref{fig:levol},
\ref{fig:tevol} \& \ref{fig:ltevol}), with successively smaller
clusters reading from left-to-right and then top-to-bottom as
numbered.
The plots go back to a time ($a=0.35$) when the smallest cluster has a
virial mass of $1.5\times10^{13}h^{-1}$\Msun, corresponding to 600
particles of each species.
Also shown is $M_{500}$, the mass enclosed by a sphere within which
the mean density is 500 times $\rho_{c0}(1+z)^3$ where $\rho_{c0}$ is
the critical density at $z=0$.

It is apparent from this figure that there are times when the mass of
a cluster increases smoothly and other times when it undergoes sudden
jumps; we designate these as minor and major mergers, respectively.
Although the division between the two is somewhat arbitrary, it is a
useful one, as we will show below that X-ray cluster luminosity
evolves quite differently in the two regimes.  For the purposes of
this paper we define the onset of a major merger as occurring when the
jump in $\log_{10}M_{\rm{vir}}$ between one output time and the next
is 0.08 or more (i.e.~a factor of 1.2).  Visual inspection of the
luminosity and temperature evolution suggests that the effects of a
major merger last for three successive output times, approximately
$1.4\times10^9\times a$\,yr.  This roughly corresponds to the
dynamical time of the clusters, although it does not scale in quite
the same manner ($\propto a$ instead of $\propto a^{3/2}$).  The jump
criterion of 0.08 is chosen such that, averaged over all the clusters,
about half of the mass is accreted during major and half during minor
mergers.  On average, about 25--30 per cent of the cluster outputs
correspond to periods of major merger activity.

From Fig.~\ref{fig:mevol}, it can be seen that clusters exhibit a wide
variety of formation histories.  Some (e.g.~clusters 9 or 16) aquire
most of their mass in major mergers, while others (clusters 2 and 20)
undergo no major mergers at all.  The percentage increase of
logarithmic mass during major mergers is show in the figure and
summarized as a histogram in Table~\ref{tab:majmerge}.  \input
majmerge.tab

It is common in the literature \citep{LaC94,NFW97} to tacitly assume
that cluster properties are fixed at the time of formation and remain
unchanged since then.  However, only in a minority of cases is it
possible to assign a particular `formation time', associated with a
major merger, to the cluster.  An alternative approach has been
suggested by \citet{WBP02} who fit models of the form
$M(a)=M_0\,e^{2a_{\rm{f}}(1-\frac{1}{a})}$ to the data.  The point
where the slope of the relation equals 2 defines a characteristic
formation epoch for each cluster, as listed in Table~\ref{tab:clus}.
We shall investigate later, in Section~4, whether the X-ray properties
of clusters are correlated with their formation time.

The growth in $M_{\rm{vir}}$ is relatively monotonic although there is
sometimes a slight decline after a merger.  The rises in $M_{500}$
occur later and are often followed by a decline as the central regions
of a cluster settle down after the merger.  For this reason, it may be
thought that $M_{\rm{vir}}$ is a more useful measure of the mass of a
cluster.  However, it will be shown in Section~4 that the X-ray
properties are more closely correlated with $M_{500}$, because the
bulk of the X-ray emission originates in this central region.

\subsection{Long-term trends in X-ray properties}

\begin{figure*}
\centerline{\psfig{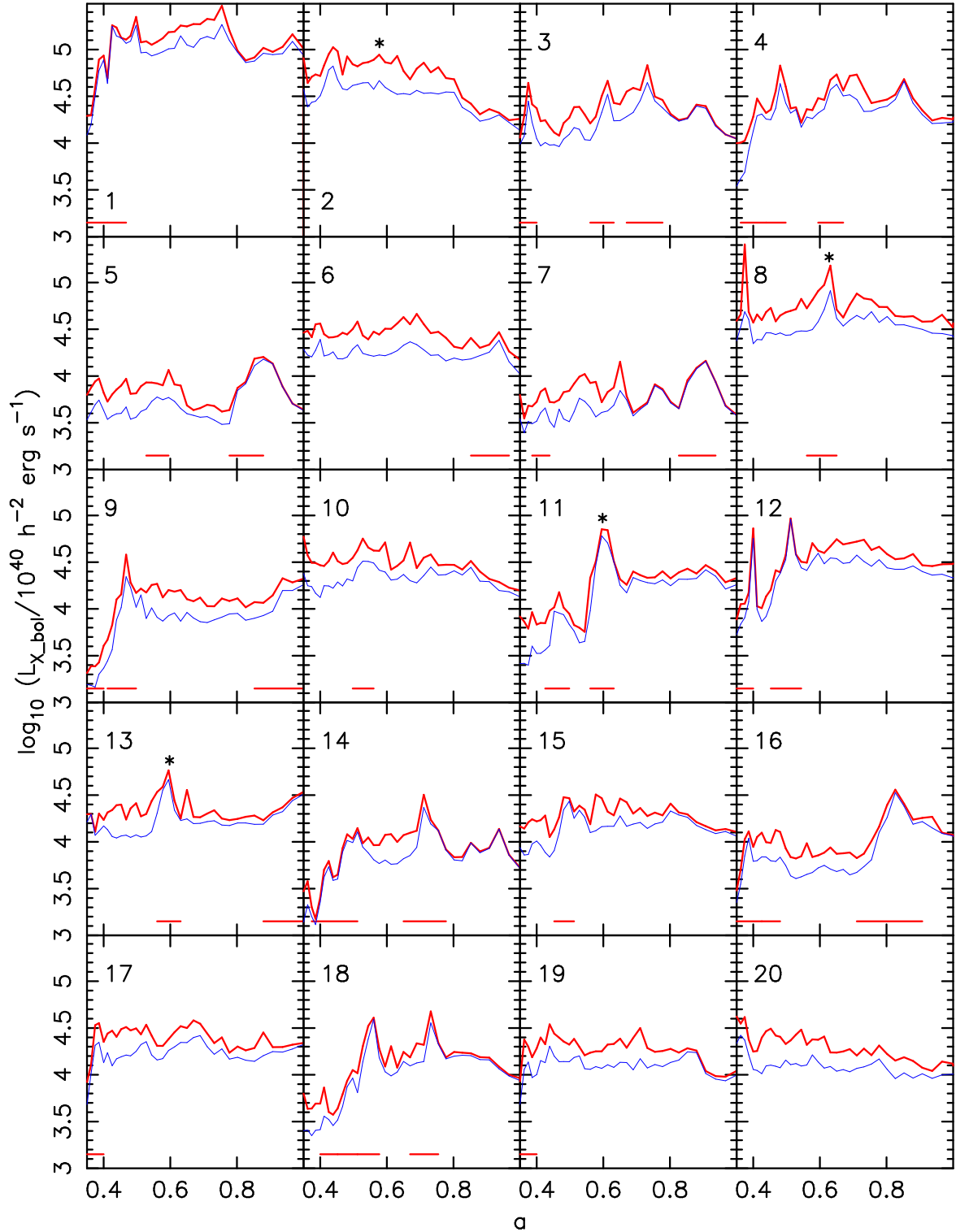}}
\caption{Bolometric luminosity versus expansion factor for the
  clusters.  The thick line shows the total emission; the thin line
  shows cooling-flow corrected emission. The asterisks indicate events
  which will be investigated in detail below. The lines along the
  bottom of each box indicate the times during which the cluster is
  deemed to be undergoing a major merger.}
\label{fig:levol}
\end{figure*}

\begin{figure*}
\centerline{\psfig{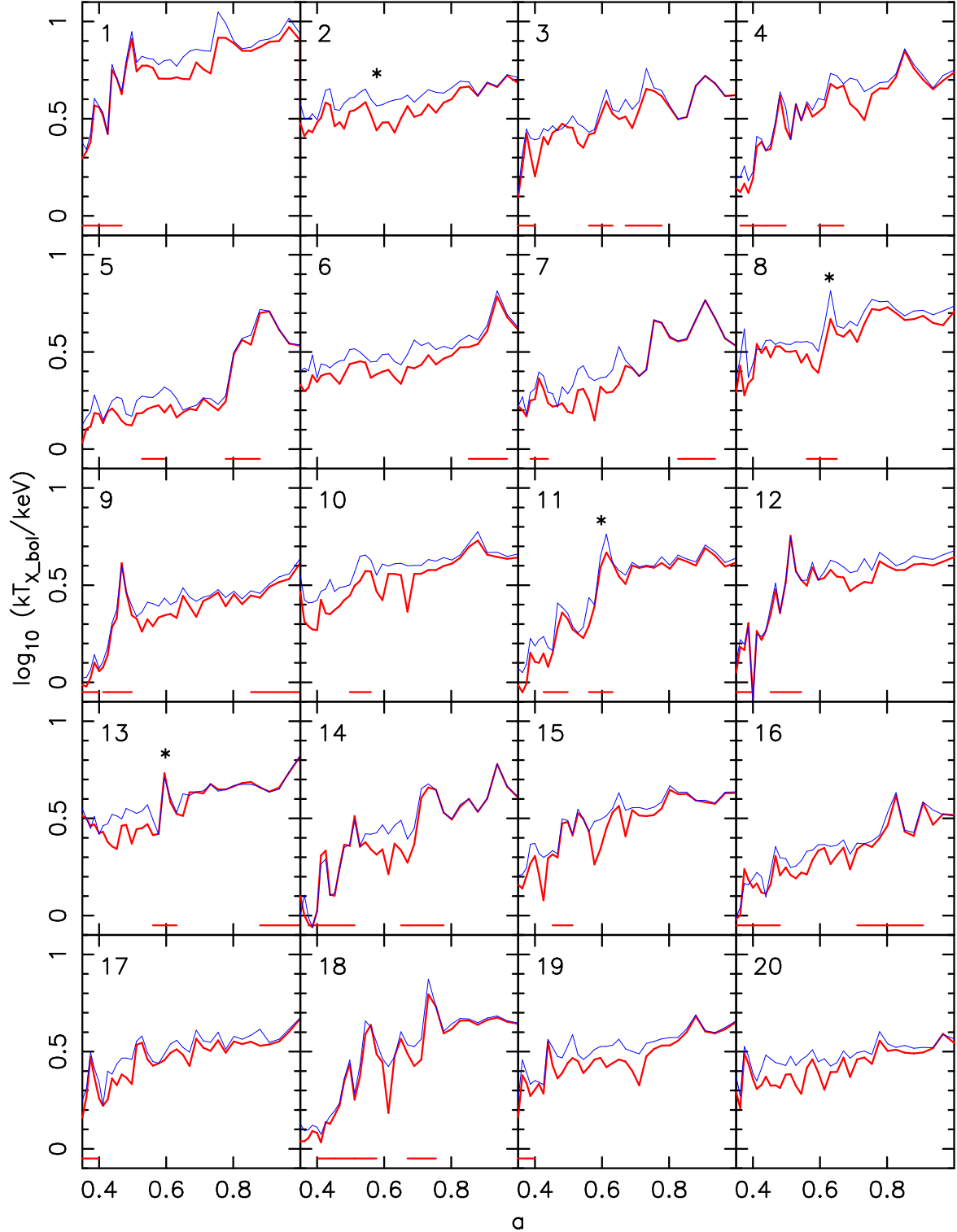}}
\caption{Bolometric emission-weighted temperature versus expansion
  factor for the clusters.  The thick line shows temperature
  calculated using the total emission; the thin line using
  cooling-flow corrected emission. The asterisks indicate events which
  will be investigated in detail below. The lines along the bottom of
  each box indicate the times during which the cluster is deemed to be
  undergoing a major merger.}
\label{fig:tevol}
\end{figure*}

Figs~\ref{fig:levol} \& \ref{fig:tevol} show the time-development of
the bolometric luminosity and emission-weighted temperature of the
clusters.
The thick line shows the total emission and the thin line shows
cooling-flow corrected emission.

These figures show that each cluster's X-ray temperature and luminosity
are undergoing continual fluctuations associated with both major and
minor mergers.  These tend to obscure the long-term trends, and so we
have defined smoothed versions of these plots in which the average
profiles are defined as the median from each set of 5 successive
output times.  

\begin{figure*}
\centerline{\psfig{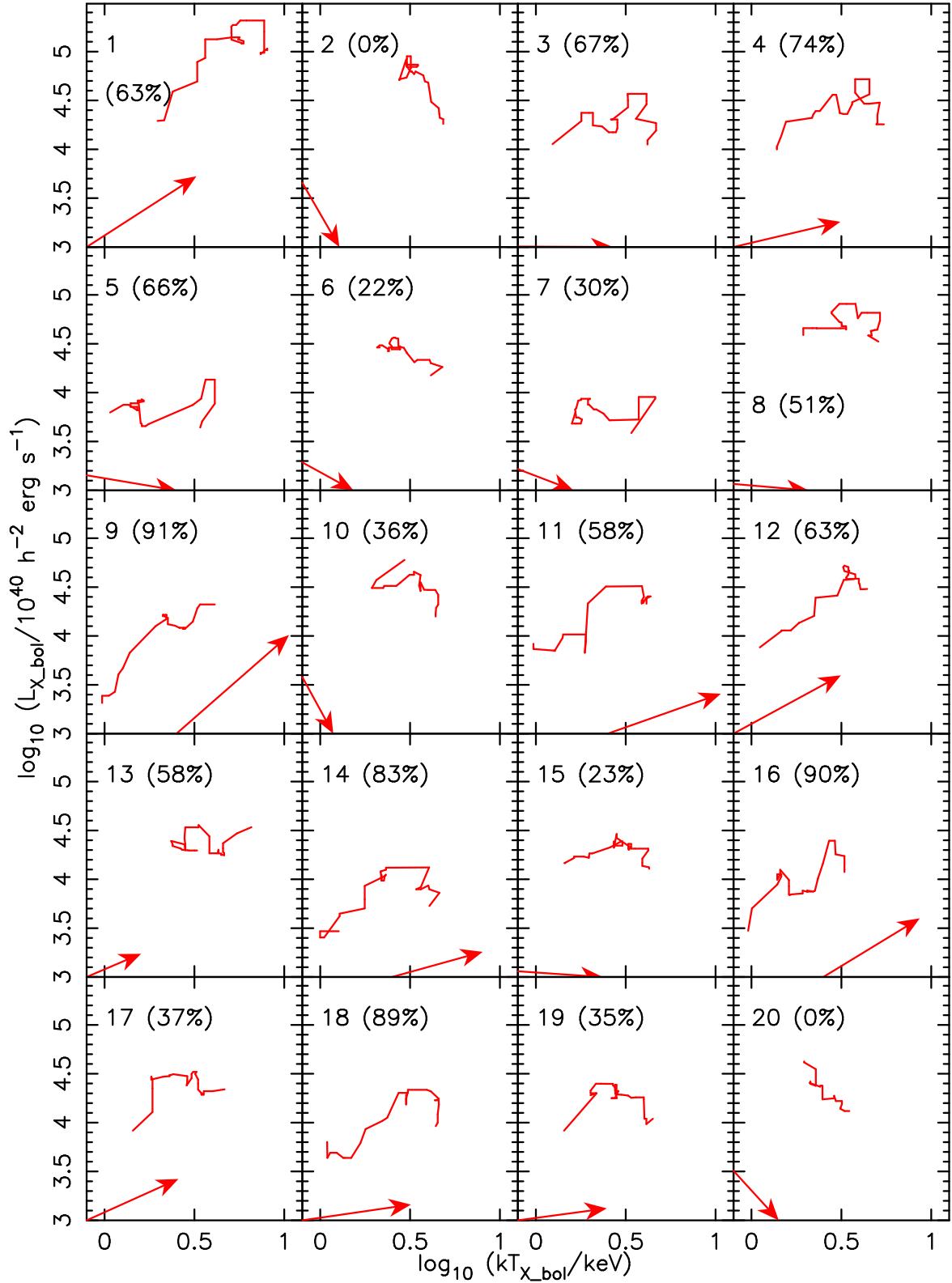}}
\caption{The time-development of the smoothed bolometric luminosity
  versus emission-weighted temperature for the clusters. The arrows
  show the overall drift of the cluster in the $\Lx-\Tx$ plane. The
  bracketed numbers indicate the fractional logarithmic mass accreted
  during major mergers.}
\label{fig:ltevol}
\end{figure*}

Fig.~\ref{fig:ltevol} shows the trajectory of the smoothed X-ray
properties for each cluster in the luminosity-temperature plane.
Also displayed are the fractional logarithmic mass accreted during
major mergers, and vectors indicating the overall drift of the cluster
across the $\Lx-\Tx$ plane.  It can be seen that the clusters that
accrete only a small amount of mass during major mergers tend to move
down and to the right (i.e.~they get less luminous with time), whereas
those that accrete most of their mass in this way move up and to the
right (i.e.~they get more luminous).  This is confirmed by the
statistics in Table~\ref{tab:majmerge} which show the average
direction of motion across the \Lx-\Tx\ plane in four bins
corresponding to different fractional logarithmic mass accreted during
major mergers.

For the majority of their lives, between major mergers, clusters grow
dimmer and heat up (on average increasing their temperature by 0.20 in
the log with a gradient in the $\log\Lx$-$\log\Tx$ plane of -1.46) so
that the normalisation of the \Lx-\Tx\ relation decreases with time.
During major mergers, clusters get brighter and hotter (on average
increasing their temperature by 0.26 in the log with a gradient of
1.54).  If we allow for the time the mergers take then this motion is
roughly parallel to the mean relation (\Lx$\propto\Tx^{2.8}$).  This
would suggest the merger history of clusters does not induce
significant scatter in the relation.  Thus we can think of the mean
\Lx-\Tx\ relation as gradually drifting to lower normalization as time
increases. Major mergers do not affect this drift but shift individual
clusters along the relation to higher temperature and luminosity.

Simple scaling arguments suggest that $L\propto M\rho T^{1/2}$ where
$M$ is the cluster mass, $\rho$ the density and $T$ the temperature.
(Note that the dominant contribution to the flux comes from the region
where $\rho\propto r^{-3/2}$ so this defines the characteristic size
to use in the scaling relation.)  For the population as a whole one
would expect the luminosity to decrease with time, at fixed mass, as
the characteristic density decreases roughtly in proportion to the
mean density of the Universe.  However, for individual clusters, the
increase in mass more than makes up for this.  However, as we have
seen above, the increase in luminosity is not smooth.  During major
mergers the mass increases abruptly.  There will be an influx of cool,
low-entropy gas and much of this will reach the core, causing a
long-term increase in luminosity.  Between major mergers the mass
continues to increase but the influx of low-entropy gas is
insufficient to replenish gas lost by radiative cooling and the core
density and luminosity decrease.  This will be looked at in more
detail in a future paper.

\subsection{Short-term fluctuations in the X-ray properties}

We next turn from the long-term behaviour of the smoothed temperature
and luminosity profiles to look at the fluctuations during mergers.
These exhibit a variety of behaviours that we shall illustrate with
specific examples drawn from our simulations.  Figs~\ref{fig:map1},
\ref{fig:map2}, \ref{fig:map3} \& \ref{fig:map4} show maps of several
example clusters at different expansion factors as indicated in the
captions. The maps are generated in the same manner as in
\citet*{OKT03}. The half-width of each map is equal to the virial
radius and the emission is projected along the line-of-sight on each
side of the clusters to a depth equal to twice the virial radius.  The
contours show X-ray surface brightness and are separated by 0.2 dex;
the colours indicate emission-weighted temperature; and the arrows
show the mass-weighted bulk flow, normalized to the highest velocity
grid point.  The corresponding evolution in the luminosity-temperature
plane is shown in Figs~\ref{fig:LTevol1}, \ref{fig:LTevol2},
\ref{fig:LTevol3} \& \ref{fig:LTevol4}; in order to show the relative
size of fluctuations, these figures all have the same dynamic range.

\subsubsection{Slow accretion}

Even when not undergoing major mergers, clusters still show
fluctuations in their luminosity and temperature evolution.  However,
these are reduced if emission from within 50$\,h^{-1}$kpc of the
cluster core is excluded (see Table~\ref{tab:delta}).
What is happening is that the clusters are growing by accretion of
small subclumps and the fluctuations are due to emission from
low-entropy gas that makes its way to the cluster core and cools
rapidly.  This sometimes causes a temporary decrease in the
temperature of the cluster core, such as the two dips in the
temperature of Cluster~10 at expansion factors of 0.60 and 0.67.

\begin{table}
\caption{Root-mean-square change in the log of the bolometric
  luminosity and temperature between one output time and the next,
  excluding times during major mergers.}
\label{tab:delta}
\begin{center}
\begin{tabular}{llc}
Property& Total& Cooling-flow corrected\\\hline
Luminosity& 0.120& 0.090 \\
Temperature& 0.069& 0.059 \\
\end{tabular}
\end{center}
\end{table}

\begin{figure*}
\centerline{\psfig{file=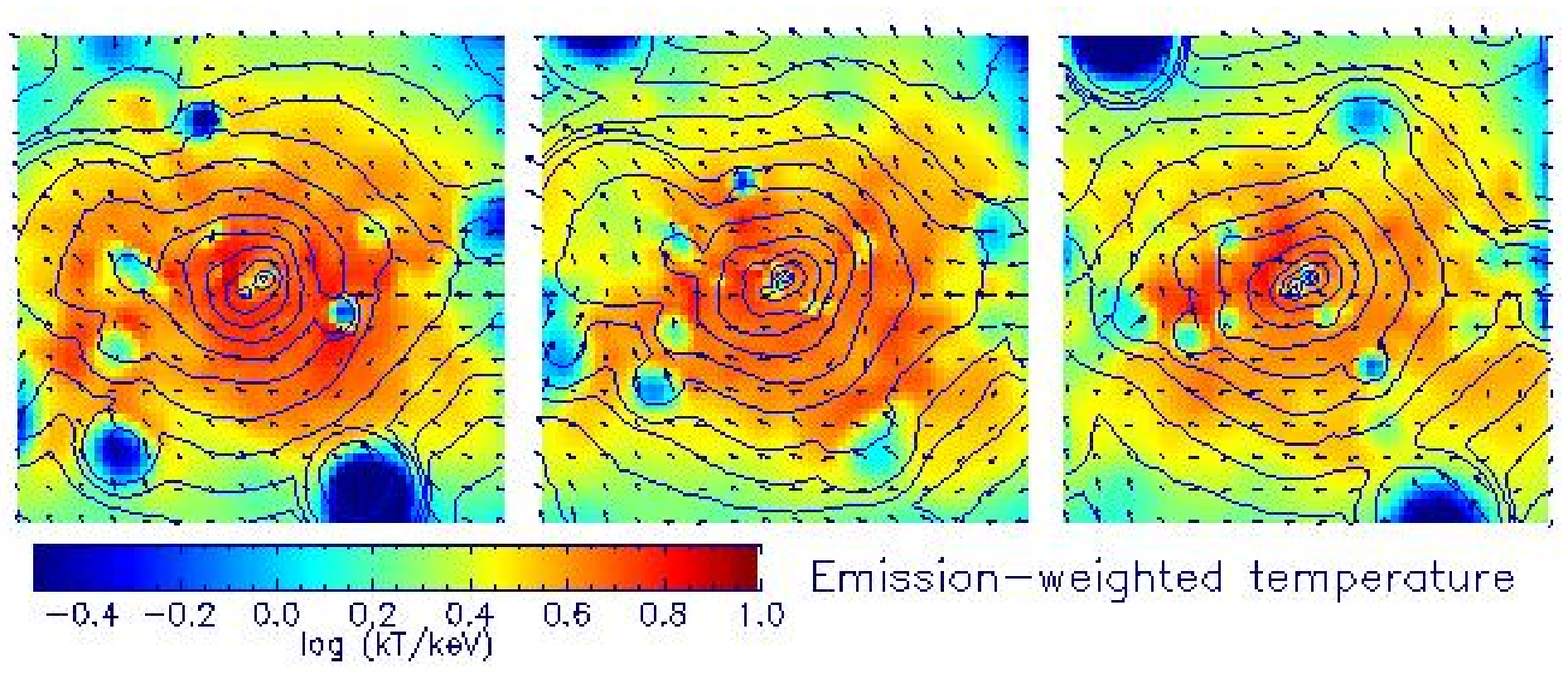,width=15.5cm}}
\caption{Surface-brightness and emission-weighted temperature map for
  Cluster~2, as described in the text. The panels are at $a=0.544$,
  0.560 \& 0.577 so there is about $2.6 \times 10^8$~yr between
  panels. The virial radius increases from $0.84 \ h^{-1}$~Mpc at the
  first panel to $0.88 \ h^{-1}$~Mpc at the third panel.}
\label{fig:map1}
\end{figure*}
\begin{figure}
\centerline{\psfig{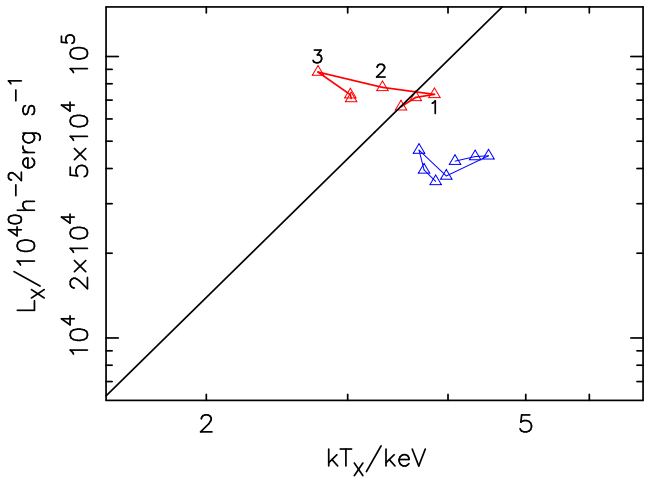}}
\caption{Evolution in the \Lx-\Tx\ plane of Cluster~2 over the time
  shown in Fig.~\ref{fig:map1}. The panels in Fig.~\ref{fig:map1}
  correspond to the numbers 1--3 here.  The thick line plots
  properties for the whole cluster and the thin line plots
  cooling-flow corrected \Lx\ and \Tx.  The straight solid line shows
  a power-law relation of the form $\Lx\propto \Tx^{2.8}$.}
\label{fig:LTevol1}
\end{figure}
Cluster~2, the second most-massive cluster in our sample, has accreted
all its mass during minor mergers.  Fig.~\ref{fig:map1} shows
snapshots at $a \approx 0.56$ in which the infall of small subclumps
is clearly visible.  The substructure statistic is less than 0.1
throughout the evolution of the cluster, indicating that the subclumps
are too small to significantly displace the centre of mass.  Even so,
the accretion gives rise to small-scale scatter as seen in
Figs~\ref{fig:levol}, \ref{fig:tevol} \& \ref{fig:LTevol1}.
At any particular time, most clusters show this kind of low-level
activity similar to the minor mergers investigated by \citet{MBL03}.

\subsubsection{Single-peaked major mergers}

The most-common behaviour seen in major mergers is a temporary upward
fluctuation in luminosity that is usually accompanied by an increase
in temperature (the residuals of \Lx\ and \Tx\ have a Spearman rank
coefficient of 0.48 which corresponds to a correlation with a
significance of over 99.9 per cent).  Many examples of this can be
detected in Figs~\ref{fig:levol} \& \ref{fig:tevol}.  These
fluctuations are not restricted to the core and correspond to more
violent mergers that boost the luminosity of the whole cluster through
compression of the intracluster medium.

\begin{figure*}
\centerline{\psfig{file=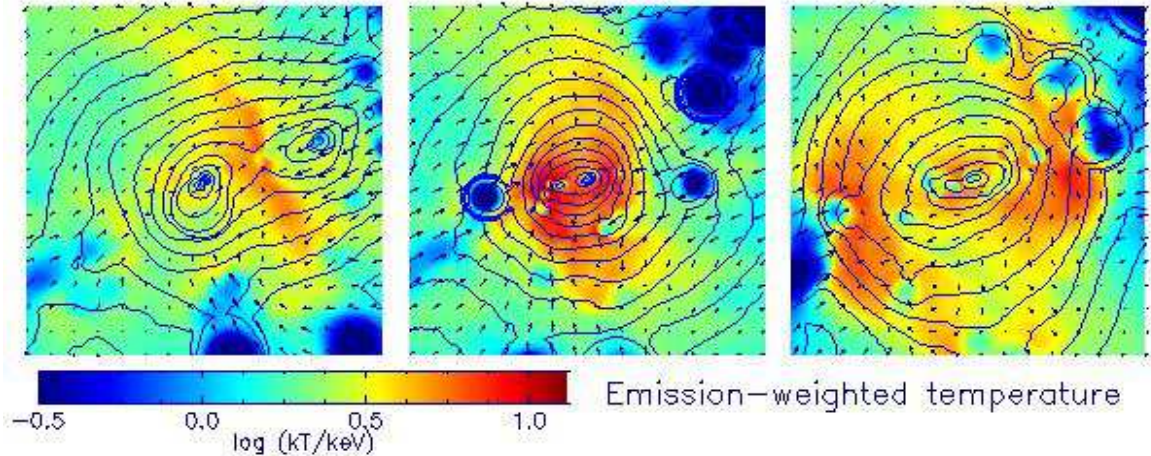,width=15.5cm}}
\caption{Surface-brightness and emission-weighted temperature map for
  Cluster~8, as described in the text. The panels are at $a=0.595$,
  0.631 \& 0.670 so there is about $5.8 \times 10^8$~yr between
  panels. The virial radius increases from $0.68 \ h^{-1}$~Mpc at the
  first panel to $0.76 \ h^{-1}$~Mpc at the third panel.}
\label{fig:map2}
\end{figure*}
\begin{figure}
\centerline{\psfig{file=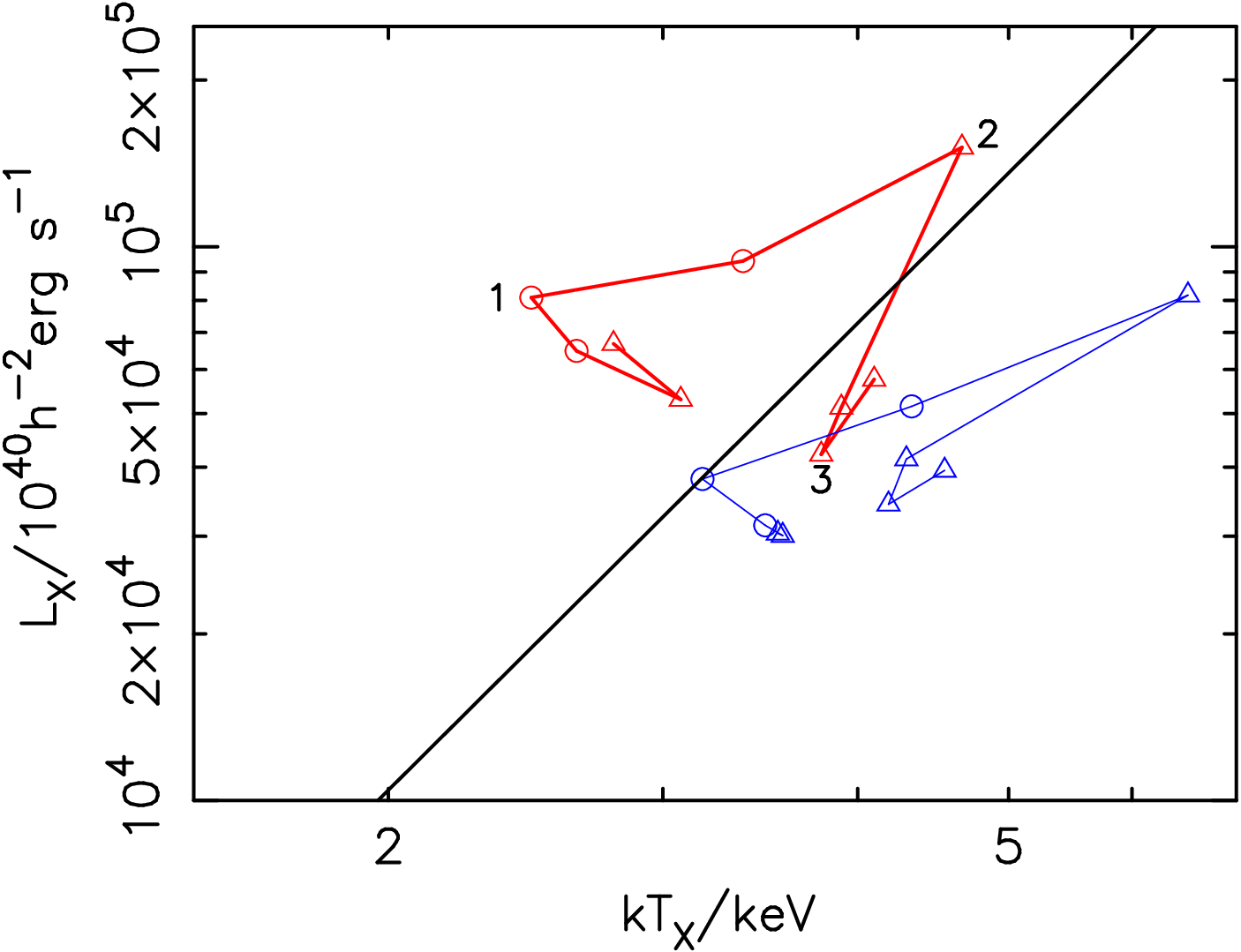,width=8.5cm,angle=270}}
\caption{Evolution in the \Lx-\Tx\ plane of Cluster~8 when it
  undergoes the merger shown in Fig.~\ref{fig:map2}. The panels in
  Fig.~\ref{fig:map2} correspond to the numbers 1--3 here.  The
  circles correspond to times when the cluster has a substructure
  statistic greater than 0.1; other times are marked with triangles.
  The thick line plots properties for the whole cluster and the thin
  line plots cooling-flow corrected \Lx\ and \Tx.  The straight solid
  line shows a power-law relation of the form $\Lx\propto \Tx^{2.8}$.}
\label{fig:LTevol2}
\end{figure}
Figs.~\ref{fig:map2} \& \ref{fig:LTevol2} show an example from
Cluster~8 at $a \approx 0.63$.  This is caused by an approximately
equal-mass merger (of clusters with masses
$1.82\times10^{14}\,h^{-1}\,\Msun$ and
$1.41\times10^{14}\,h^{-1}\,\Msun$) that produces a planar compression
front perpendicular to the direction of the merger.  The planar nature
of the front can be more or less obvious than that shown here,
depending upon the impact parameter of the collision, the degree of
substructure and the viewing angle.

\subsubsection{Double-peaked major mergers} 

Sometimes a merger will be associated with a double peak in the
luminosity, the first associated with an increase in temperature and
the second a decrease. This happens when the cores of the main cluster
and the in-falling subcluster do not merge directly but orbit about
each other for a while before doing so.  Unfortunately the time
resolution of our outputs is insufficient to determine whether this is
the generic behaviour.

\begin{figure*}
\centerline{\psfig{file=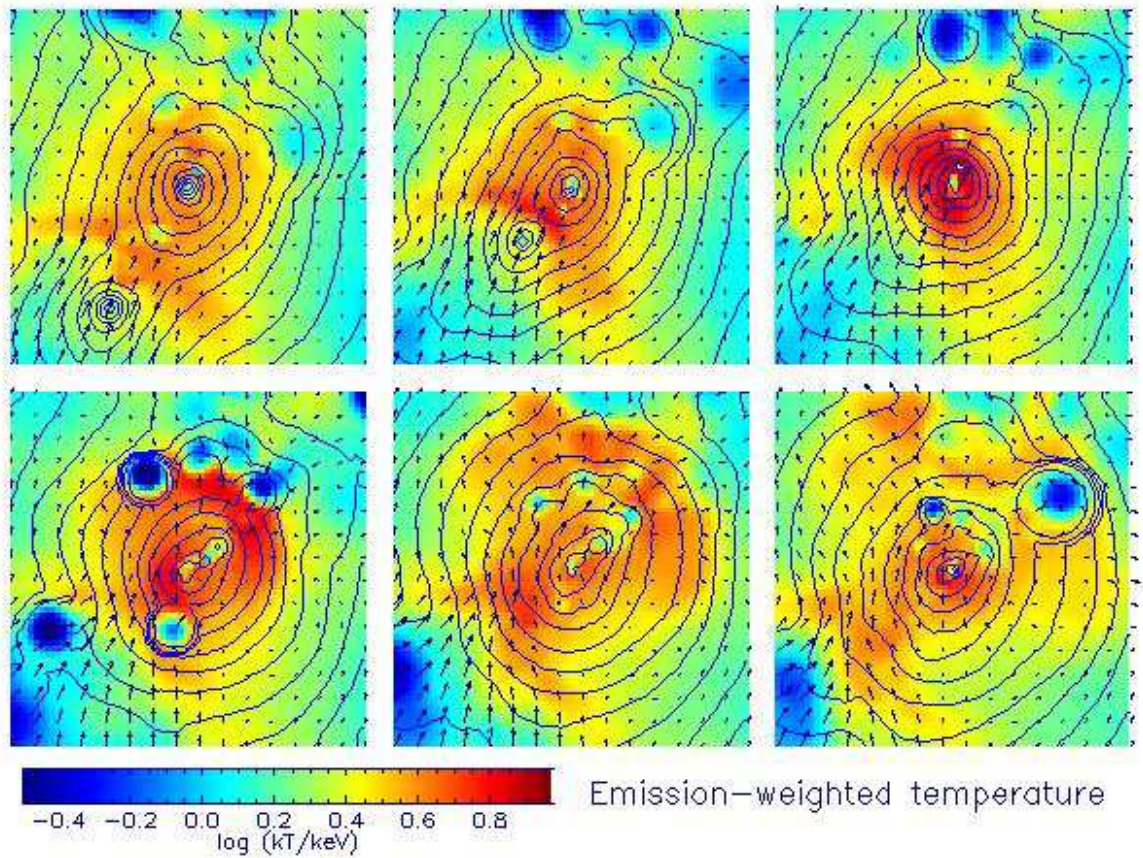,width=15.5cm}}
\caption{Surface-brightness and emission-weighted temperature map for
  Cluster~13, as described in the text. The first panel is at
  $a=0.560$ and the last is at $a=0.650$ so there is about $2.8 \times
  10^8$~yr between each panel. The virial radius increases from $0.72
  \ h^{-1}$~Mpc at the first panel to $0.95 \ h^{-1}$~Mpc at the sixth
  panel.}
\label{fig:map3}
\end{figure*}
\begin{figure}
\centerline{\psfig{file=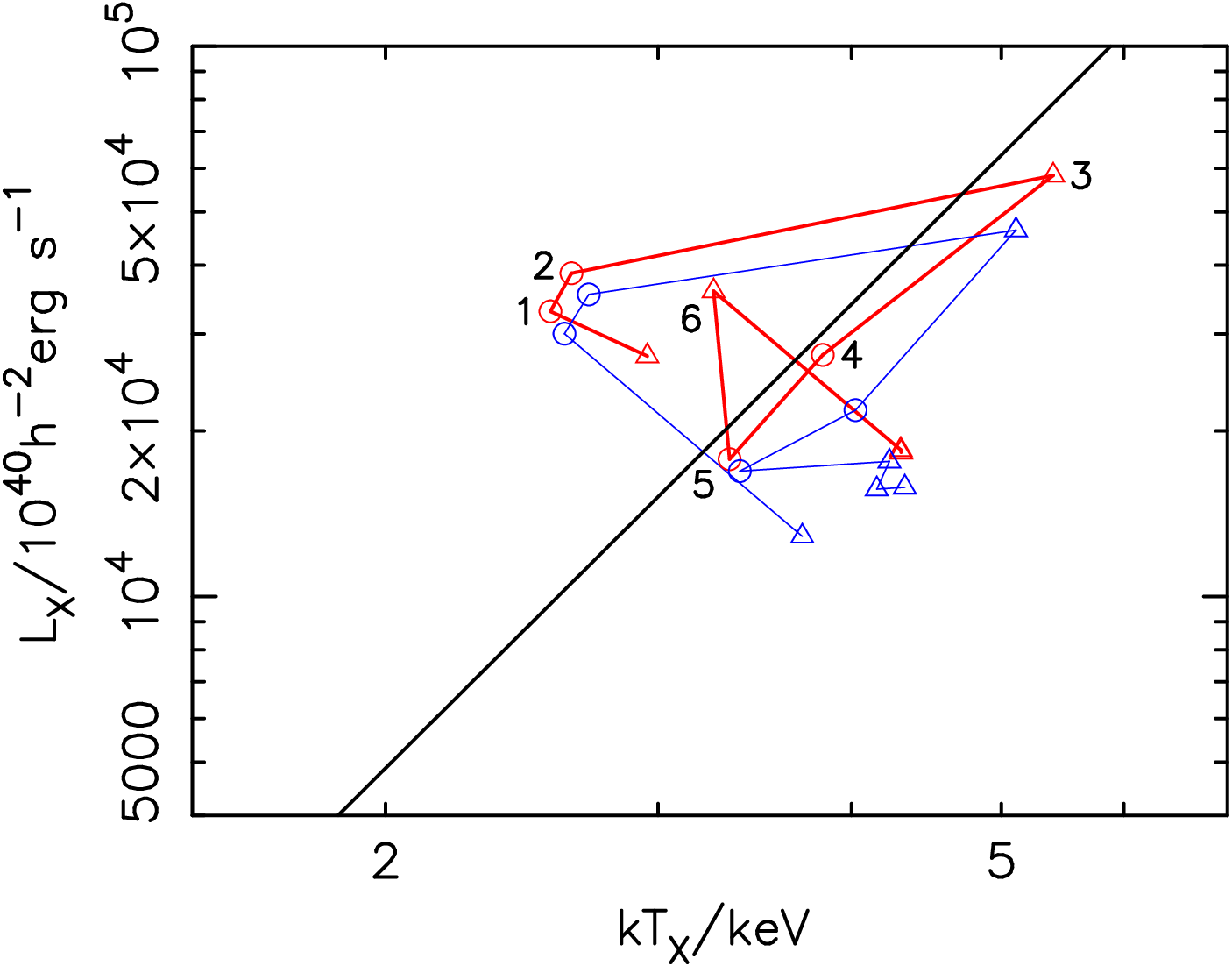,width=8.5cm,angle=270}}
\caption{Evolution in the \Lx-\Tx\ plane of Cluster~13 when it
  undergoes the merger shown in Fig.~\ref{fig:map3}. The panels in
  Fig.~\ref{fig:map4} correspond to the numbers 1--6 here.  The
  circles correspond to times when the cluster has a substructure
  statistic greater than 0.1; other times are marked with triangles.
  The thick line plots properties for the whole cluster and the thin
  line plots cooling-flow corrected \Lx\ and \Tx.  The straight solid
  line shows a power-law relation of the form $\Lx\propto \Tx^{2.8}$.}
\label{fig:LTevol3}
\end{figure}
Fig.~\ref{fig:map3} shows a major merger of this type in Cluster~13 at
$a \approx 0.6$. In the first two panels the subclump approaches and
in the third panel the clusters collide and the temperature and
luminosity peaks.  At the time of maximum luminosity, however, the
surface brightness contours are relatively round and there is little
evidence from the surface-brightness alone that a merger is taking
place (the asymmetry that is visible in the figure is from in-falling
matter that does not contribute significantly to the luminosity and
temperature variation).  The cores have not lost all relative momentum
however and so they reseparate and do not fully merge until the final
panel. This brings about a second peak in the luminosity which is only
observed in the core (see Fig.~\ref{fig:LTevol3}).  Since this is
relatively cool, the X-ray temperature drops at this point before
stabilizing at a higher value.  Note that, at the point the cores
merge, the cluster again looks fairly spherical and relaxed.

This is the behaviour observed in the toy models of \citet{RiT02}. The
largest peak in the luminosity in these simulations was when the
clusters first collided.  Then the luminosity dropped as the core
expanded again and then increased to a stable level as the core
finally settled.  This behaviour was more pronounced in off-axis
collisions where the cores missed each other on the first pass.

\subsubsection{Major mergers leading to a permanent increase in luminosity}

Finally, there are major mergers that cause a permanent jump in
luminosity and temperature after the initial fluctuation associated
with the merger has died away. Half of the mergers involve a jump in
the smoothed \Lx\ (i.e.~a permanent jump in \Lx) of over a factor of 2
and half of the mergers involve a jump in the smoothed \Tx\ of more
than a factor of 1.4.

\begin{figure*}
\centerline{\psfig{file=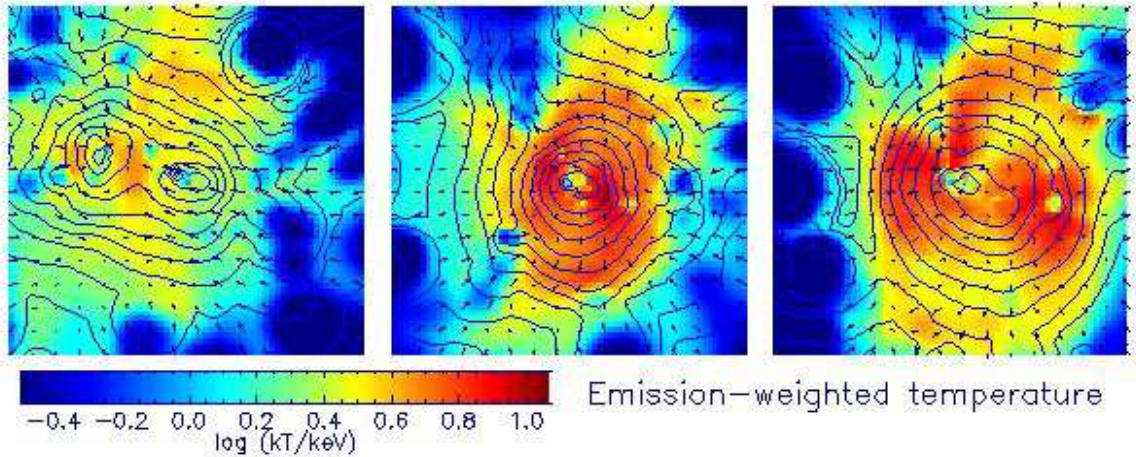,width=15.5cm}}
\caption{Surface-brightness and emission-weighted temperature map for
  Cluster~11, as described in the text. The panels are at $a=0.577$,
  0.613 \& 0.631 so there is about $5.5 \times 10^8$~yr between the
  first 2 panels and about $2.9 \times 10^8$~yr between the second and
  third panels. The virial radius increases from $0.89 \ h^{-1}$~Mpc
  at the first panel to $1.04 \ h^{-1}$~Mpc at the third panel.}
\label{fig:map4}
\end{figure*}
\begin{figure}
\centerline{\psfig{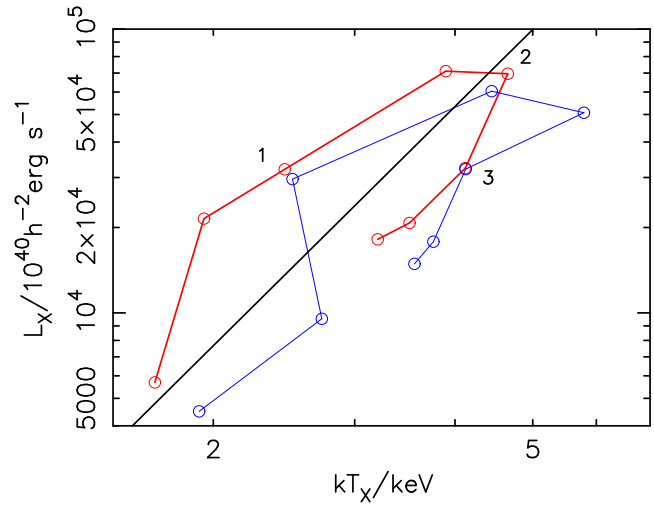}}
\caption{Evolution in the \Lx-\Tx\ plane of Cluster~11 when it
  undergoes the merger shown in Fig.~\ref{fig:map4}. The panels in
  Fig.~\ref{fig:map4} correspond to the numbers 1--3 here.  The thick
  line plots properties for the whole cluster and the thin line plots
  cooling-flow corrected \Lx\ and \Tx.  The straight solid line shows
  a power-law relation of the form $\Lx\propto \Tx^{2.8}$.}
\label{fig:LTevol4}
\end{figure}
Fig.~\ref{fig:map4} shows images before, during and after one such
merger which Cluster~11 undergoes at $a\approx0.6$.  There is a small
amount of hot, compressed gas as the subclump approaches but not so
obviously planar as in some other mergers. At the point of merging the
cluster looks fairly relaxed in this projection although the
substructure statistic is greater than 0.1 throughout.

Fig.~\ref{fig:LTevol4} shows that \Lx\ and \Tx\
increase together approximately parallel to the \Lx-\Tx\ relation in
an elongated, clockwise ellipse.  This is because the luminosity
increases when the clump crosses $R_{\rm{vir}}$, before it has had a
chance to interact strongly with the ICM and raise the temperature of
the gas.  A similar effect is seen in Fig.~\ref{fig:LTevol2} and the
first (main) merger in Fig.~\ref{fig:LTevol3}

\section{Scatter in the X-ray scaling relations}

To a greater or lesser degree all of the X-ray scaling relations show
scatter both in observations and simulations.  Some of this will be
due to observational errors and resolution effects.  However, as
clusters have differing merger histories and are not self-similar, it
is apparent that some, perhaps most, of this scatter is physical.  As
Figs~\ref{fig:levol} \& \ref{fig:tevol} show, much of the scatter in
luminosity and temperature comes from gas within $50\,h^{-1}$\,kpc of
the cluster centre.  For this reason, we omit the central gas when
plotting the scaling relations.

\begin{figure}
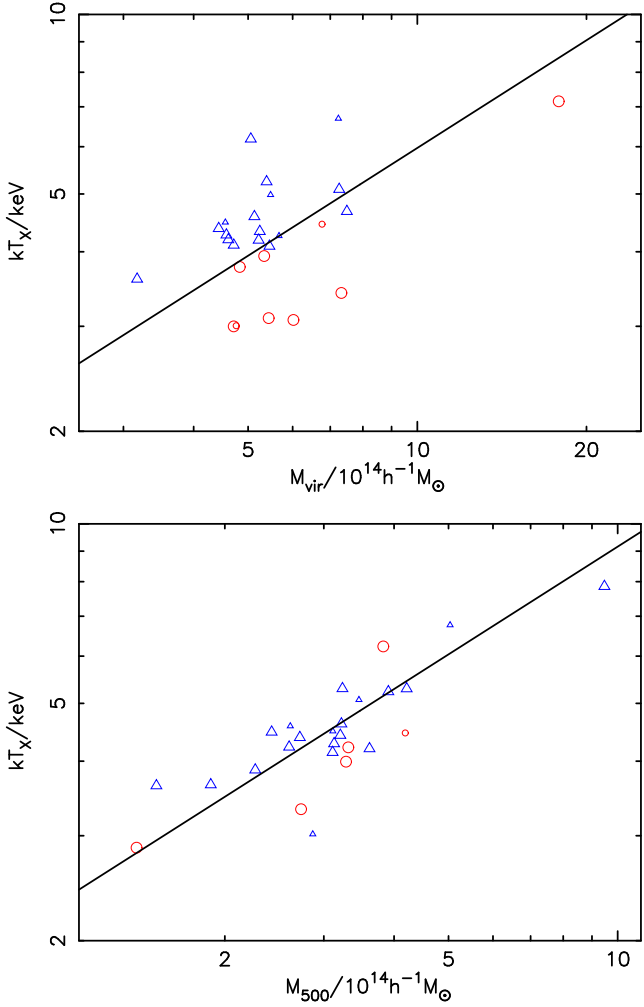

\centerline{\psfig{file=txm_Aboth.ps,width=8.5cm,angle=270}}
\vspace{0.2cm}
\centerline{\psfig{file=txm_Cboth.ps,width=8.5cm,angle=270}}
\caption{Cooling-flow corrected soft-band X-ray temperature versus
  mass.  In the upper panel the outer radius of the cluster is $R_{\rm
  vir}$; in the lower panel it is $R_{500}$.  The solid lines are fits
  to the data of a power-law relation $\Tx\propto M^{0.60}$.  The
  circles represent clusters for which the substructure statistic is
  greater than 0.1; other clusters are indicated with triangles.
  Large symbols represent the resimulated clusters presented in this
  paper; small symbols are clusters from MTKP02.}
\label{fig:txm}
\end{figure}

Fig.~\ref{fig:txm} shows the current-day X-ray temperature-mass
relation within $R_{\rm vir}$ (top panel) and $R_{500}$ (bottom
panel).  The straight line is the best-fit relation to a straight line
of slope 0.60 in log space.  This slope was chosen to agree with that
from the clusters in MTKP02 which cover a much wider dynamic range in
mass.  We have investigated a wide variety of properties to try to
find the main cause of the scatter.  These include the following:
\begin{itemize}
\item The most obvious source of scatter is the fluctuations in
  temperature associated with mergers, i.e.~the difference between the
  actual temperature and the smoothed temperature.  However, at the
  present day this is of a magnitude (approximately 0.03 dex) that is
  too small to have much of an effect.  We have looked for a
  correlation in these fluctuations and deviations from the
  temperature-mass relation, and find none.
\item The fractional logarithmic mass accreted during major mergers is
  also uncorrelated to scatter.
\item \citet{WBP02} found a correlation between cluster formation time
  and the concentration of the dark matter.  We used their method to
  assign characteristic formation epochs to each cluster (see
  Table~\ref{tab:clus}) but again found no correlation with scatter.
\item The rate of increase of mass of both the cluster as a whole and
  of the core mass, averaged over different time periods before the
  present.  This was also unsuccessful.
\end{itemize}

In the end, only one of the statistics that we looked at gave a strong
correlation with scatter from the X-ray temperature-mass relation and
that was the degree of substructure.  The substructure statistic,
defined as the separation between the position of the centroid and the
dark-matter density maximum in units of the cluster radius, is listed
in Table~\ref{tab:clus}.  (Note that this statistic is a function of
the cluster radius; the values in the table are for properties
averaged within the virial radius.)  The substructure statistic has
several advantages over some of the other measures that we tried: it
is relatively simple to calculate, it depends only upon the properties
of the cluster at the present day, and it is closely related to some
observable quantity (for example \citet{SBR01} observed substructure
in $52\pm7$ per cent of clusters from the REFLEX+BCG sample).

In Fig.~\ref{fig:txm}, clusters with a substructure statistic greater
than 0.1 are plotted as circles.  It is immediately apparent from the
upper panel that a large part of the scatter is related to
substructure (formally, the distance below the line correlates to
substructure with a Spearman rank coefficient of 0.485, a significance
of more than 95 per cent).  It is interesting that most clusters with
substructure scatter low on this plot.  Some of these contain
subclumps that are falling into the cluster for the first time and
which have raised its mass without yet significantly altering the
emission-weighted temperature; others are clusters which are
undergoing a core-bounce after a merger and for which the temperature
is temporarily slightly too low.

\begin{table}
\caption{Scatter about the fits to the plots in Figs~\ref{fig:txm} to
  \ref{fig:ltx}.  This is defined as the root-mean-square deviation in
  the vertical direction in dex after allowing for the one free
  parameter in the fit.}
\label{tab:scatter}
\begin{center}
\begin{tabular}{lcc}
Relation& Scatter in $R_{\rm{vir}}$ values& Scatter in $R_{500}$ values\\\hline
$\Tx-M$& 0.088& 0.057 \\
$\Lx-M$& 0.25& 0.19 \\
$\Lx-\Tx$& 0.16& 0.15
\end{tabular}
\end{center}
\end{table}
It is apparent from the larger scatter in the upper panel of
Fig.~\ref{fig:txm}, that the virial mass cannot be accurately
determined from a cluster's X-ray properties when spatial information
is not available.  However, we can reduce the scatter considerably by
moving to a more compact radius (see Table~\ref{tab:scatter}).  This
has several advantages: the number of clusters showing substructure is
reduced, the properties are measured at radii that are more accessible
to X-ray observations, and the substructure that is measured is more
directly related to subclumps that are interacting strongly with the
intracluster medium and that will be influencing the X-ray properties.

The lower panel in Fig.~\ref{fig:txm} shows the \Tx-$M_{500}$
relation.  As can be seen, the relation is much tighter than before,
and interestingly, the scatter of high substructure clusters from the
relation is much-reduced.  The circle that lies furthest above the
solid line is Cluster~13 which has recently merged and whose
temperature is still enhanced by the associated compression of the
ICM.  This cluster was even further from the line in the original plot
(the topmost large triangle) but did not show up as having
substructure because the subclump was too close to the centre.  The
circle that lies furthest below the line in the bottom panel is
Cluster~5 which temporarily has a cooler temperature than the average
due to core expansion following a recent merger.

Note that, for most of the clusters, the mass decreases slightly in
the move from $R_{\rm{vir}}$ to $R_{500}$, but for the two left-most
clusters in the bottom panel (Clusters~7 and 14) it changes
significantly.  It can be seen from Table~\ref{tab:clus} that these
are the two clusters with the highest degree of substructure at the
end.  What has happened is that while $R_{\rm{vir}}$ encloses the
subclump causing the substructure $R_{500}$ does not.  The luminosity
of these clusters also decreases significantly and they would be
easily seen as bimodel in any X-ray observation.  For this reason, we
omit them from the scatter statistics listed in
Table~\ref{tab:scatter}.

Having eliminated the principal cause of the scatter in the \Tx-$M$
relation, we again tested to see whether any of the properties
considered earlier now correlate with the residual scatter, but with
negative results.

\begin{figure}
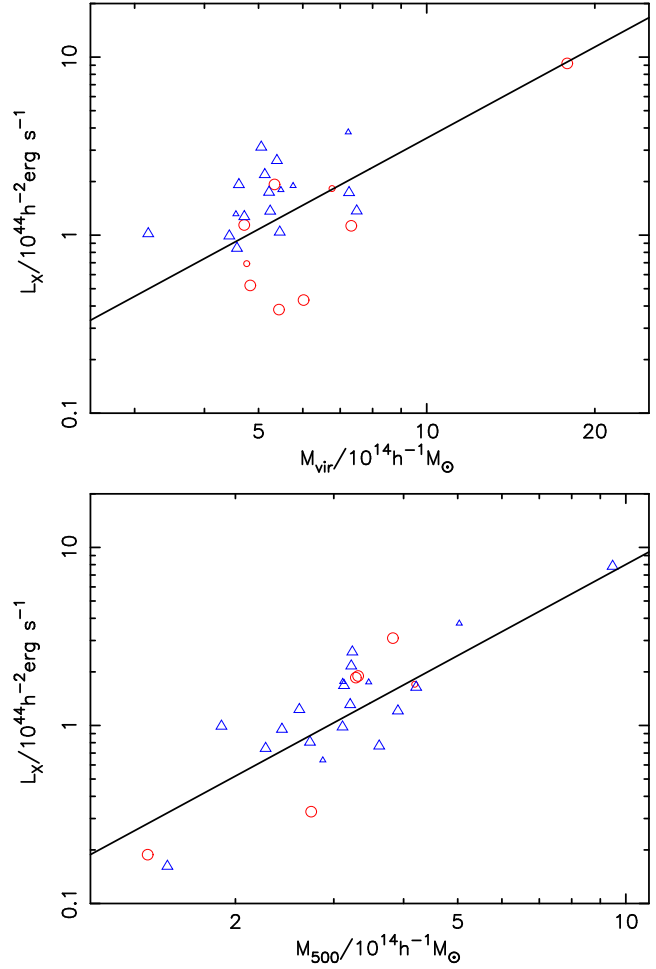

\centerline{\psfig{file=lm_Aboth.ps,width=8.5cm,angle=270}}
\vspace{0.2cm}
\centerline{\psfig{file=lm_Cboth.ps,width=8.5cm,angle=270}}
\caption{Cooling-flow corrected bolometric X-ray luminosity estimated
  from emission in the soft band versus mass.  In the upper panel the
  outer radius of the cluster is $R_{\rm vir}$; in the lower panel it
  is $R_{500}$.  The solid lines show power-law relations of the form
  $\Lx\propto M^{1.7}$.  The symbols have the same meaning as in
  Fig.~\ref{fig:txm}.}
\label{fig:lm}
\end{figure}

\begin{figure}
\centerline{\psfig{file=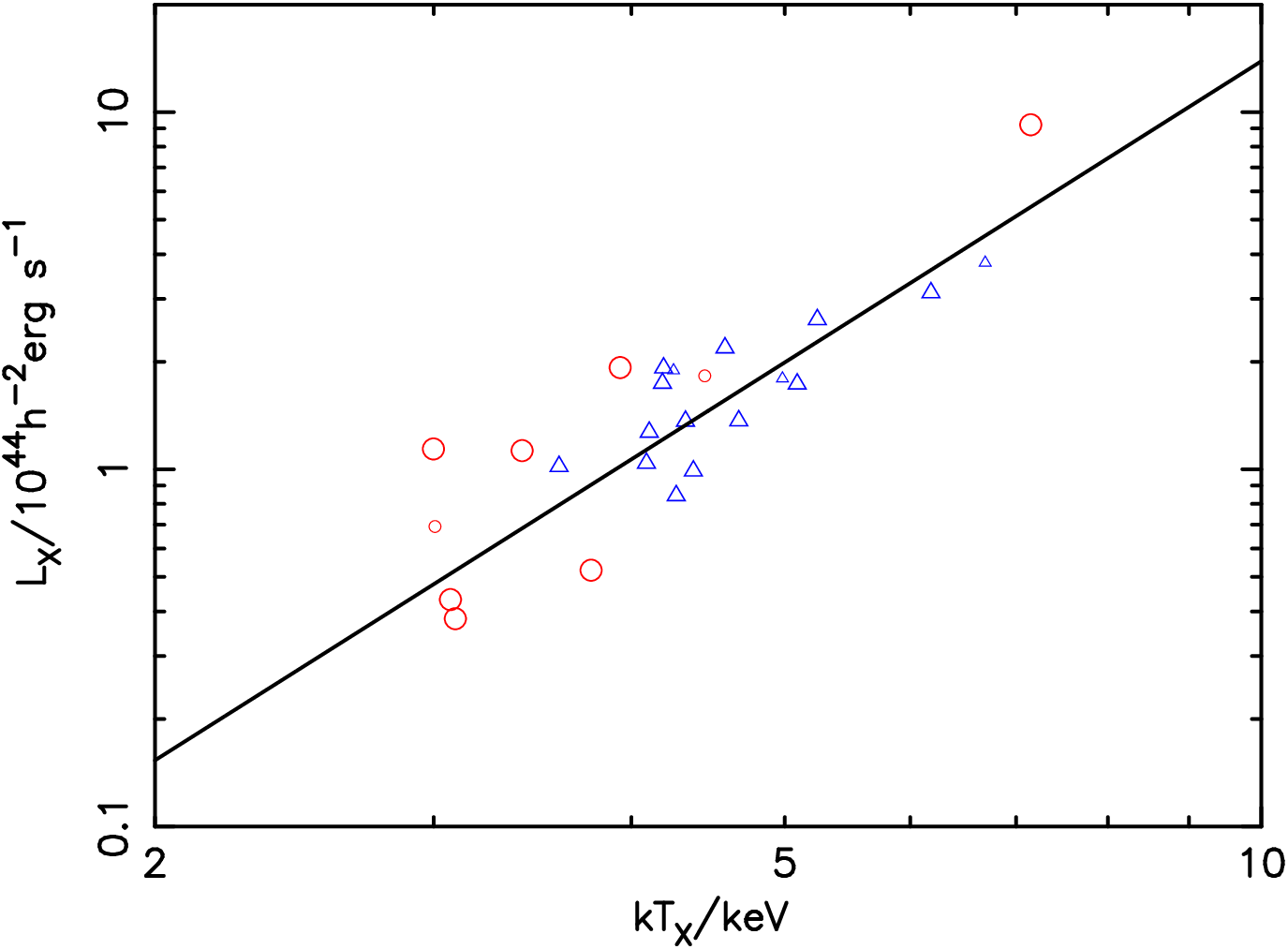,width=8.5cm,angle=270}}
\vspace{0.2cm}
\centerline{\psfig{file=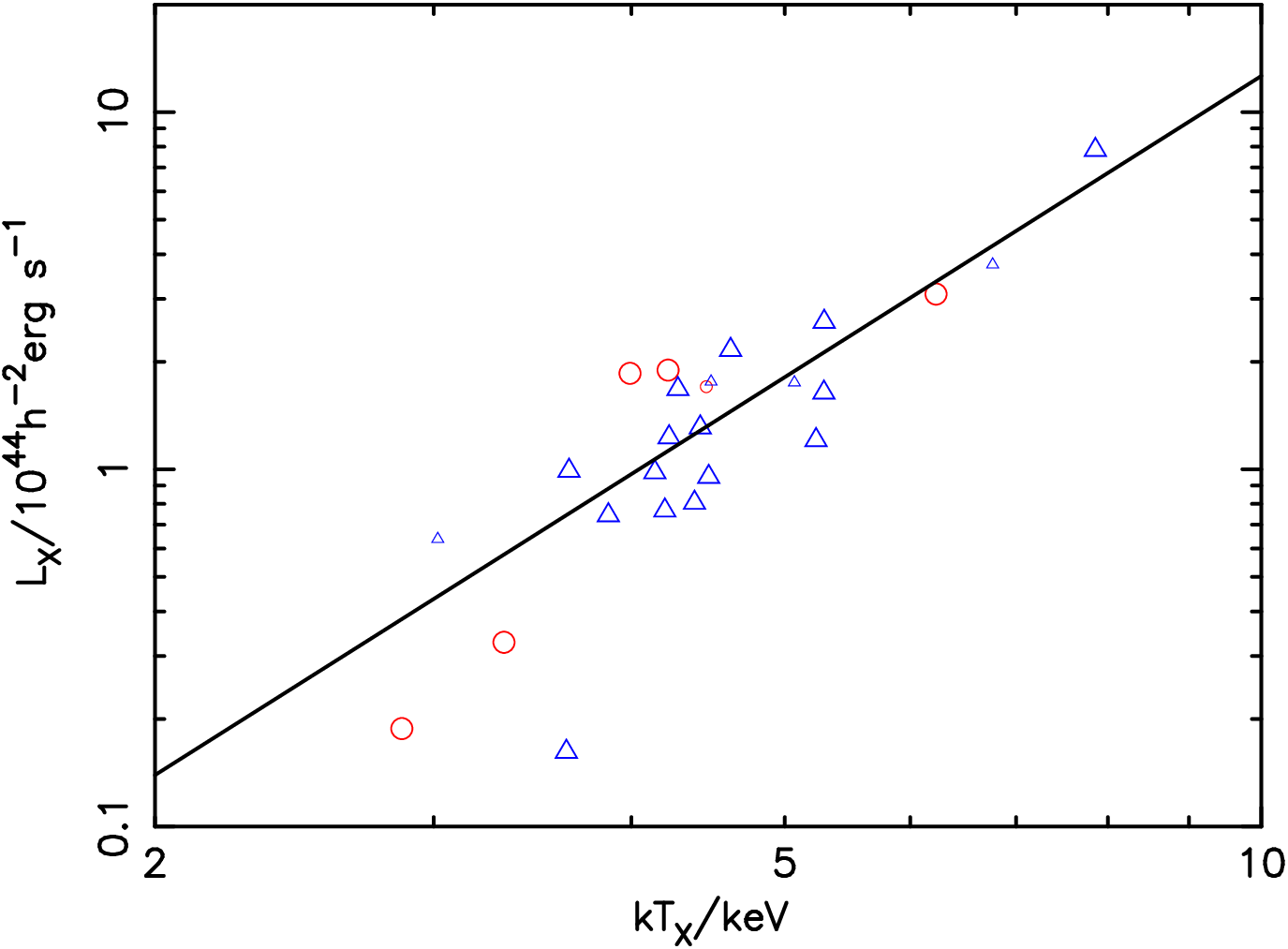,width=8.5cm,angle=270}}
\caption{Cooling-flow corrected bolometric luminosity versus X-ray
  temperature estimated from emission in the soft band.  In the upper
  panel the outer radius of the cluster is $R_{\rm vir}$; in the lower
  panel it is $R_{500}$.  The solid lines show power-law relations of
  the form $\Lx\propto \Tx^{2.8}$.  The symbols have the same meaning
  as in Fig.~\ref{fig:txm}.}
\label{fig:ltx}
\end{figure}

Figs~\ref{fig:lm} and \ref{fig:ltx} show equivalent plots for the
luminosity-mass and luminosity-temperature relations.  The former has
a similar variation with substructure as seen in the temperature-mass
relation.  The latter shows no obvious correlation of scatter with
substructure, consistent with our earlier observation that motion in
the \Lx-\Tx\ plane during mergers tends to be parallel to the mean
relation.  However, in the upper panel of Fig.~\ref{fig:ltx} there is
a concentration of clusters toward the bottom-left showing
substructure.  This reflects the fact that the sample was selected on
the basis of virial mass and contains a number of clusters for which
the merging is not complete.  Conversely, in any X-ray luminosity or
temperature-limited sample there will be a bias against high virial
mass clusters with a high degree of substructure.  This bias is
reduced when properties are measured within $R_{500}$ instead.

\section{Discussion}

In this study, the formation of 20 large clusters of galaxies has been
followed by resimulating the regions around them at high resolution.
The X-ray properties were found to vary with time, driven mainly by
the merging history of the clusters. Accretion during major mergers
tends to increase a cluster's luminosity, pushing it up the \Lx-\Tx\
relation, while between major mergers it slowly decreases, following
the movement of the \Lx-\Tx\ relation.  In addition to this long-term
variation, there are short-lived fluctuations associated with the
merger itself.

Over the period investigated clusters were said to be undergoing a
major merger about 25--30 per cent of the time.  However, to convert
this to a number that can be compared to X-ray observations is very
complex.  Firstly, our definition of a major merger was very ad-hoc
and a more detailed comparision of possible definitions with X-ray
observability would be required.  In addition, there will be strong
selection effects that favour high-redshift clusters undergoing
temporary boosts in X-ray luminosity.  A detailed investigation of
these biases will be required in order to correcctly interpret
observations of the X-ray properties of clusters at high redshift.

As a subclump crosses $R_{\rm_{vir}}$, the mass and the luminosity
increase but the temperature stays roughly constant or decreases
slightly (since the subclump will be cooler than the cluster). This
causes the cluster to move below the \Tx-$M$ and \Lx-$M$ relations and so
clusters with structure within $R_{\rm_{vir}}$ tend to be scattered
low on the plots.  For this reason, the mass within a smaller region
such as $R_{500}$ correlates more closely with X-ray properties than
does the virial mass.

As the subclump moves through the ICM of the cluster, it compresses the
gas in front of it. This causes adiabatic heating and can be observed
as a hot planar compression front.  These features have been observed,
for example by \citet{MJE03}.  Fig.~5 from their paper shows
surface-brightness contours from a {\it{Chandra}} map of Cl
J0152.7-1357 a cluster at $z = 0.833$ and shows that the cluster is
undergoing a major merger. Fig.~11 from the same paper shows the
hardness of the X-rays with lighter regions representing harder
radiation.  Given that harder X-rays will be produced by higher
temperature regions it can be seen that the gas between the merging
clusters has been heated as it is compressed, similar to our clusters
(see panel~1 of Figs.~\ref{fig:map2}, \ref{fig:map3} \&
\ref{fig:map4}).

By the time the subclump reaches $R_{500}$ the cluster will be
starting to undergo a merger boost and will start to move up in the
\Lx-$M$ and \Tx-$M$ planes back toward the mean relations.  This means
that clusters with structure will not necessarily scatter off the
\Tx-$M$ or \Lx-$M$ relations when properties within this radius are
considered.  Usually the luminosity and temperature increase together
pushing the cluster roughly parallel to the \Lx-\Tx\ relation.
However, if the merger triggers cooling within the cluster core then
the X-ray temperature decreases and the cluster will scatter
above the \Lx-\Tx\ relation and below the \Tx-$M$ relation.

As the subclump moves toward the core the compression front will get
hotter and closer to the core of both the subclump and the
cluster.  This means that it will start to ram-pressure strip the
subclump of its diffuse ICM. Therefore even if the subclump's core
survives the merger for a time, the diffuse gas will be assimilated
into the cluster by the peak of the merger.

Once the subclump has reached the core, the luminosity and temperature
boost will have reached their maximum but, unless the infalling clump
has a significant impact parameter, the cluster will be roughly
spherical and appear to have little substructure. \citet{GVS04}
observe that the luminous, $z=0.783$ cluster MS1137.5+6625 appears
spherical in the optical and in X-rays but closer inspection with
{\it{Chandra}} observations show that the cluster is not relaxed. The
mass distribution is very compact, consistent with a large amount of
recently accreted material.  Since the luminosity can be increased at
this time by up to an order of magnitude above what a truly relaxed
cluster of similar mass would be, then this could impose a bias in a
flux limited sample of clusters, particularly at high redshift, toward
clusters at the point of merging.

The effect of merger boosts in biasing the estimate of cluster
parameters from high-redshift X-ray cluster observations has been
investigated by \citet{RSR02}.

After the peak of the collapse, the core of the subclump will continue
to move past or through the core of the cluster.  This behaviour has
also been observed. {\it{Chandra}} observations of 1E0657-56
($z=0.296$) presented by \citet{MGD02} observe a `bullet' which is
surmized to be the core of a subcluster.  The subcluster has
previously passed through the core of the main cluster (some
0.1--0.2\,Gyr ago) and has had its surrounding gas removed by
ram-pressure stripping. It is unclear whether this object exceeds the
escape speed of the cluster or not, but it still shows that a subclump
can pass through the core of a cluster and that the haloes can merge
while the two cores continue on their paths.

If, as is likely, the core of the infalling subclump does not exceed
the escape speed, then it will at some later point return to the core
and merge. This will cause the luminosity of the core to increase once
more.  If the temperature profile does not change significantly when
this happens, then the increase in emission at the cooler core will
cause an emission-weighted temperature for the whole cluster to
decrease. This will move the cluster up and to the left on the \Lx-\Tx\
plane pushing the cluster above the \Lx-\Tx\ relation and so could
contribute to the scatter about the relation.  By this time the rest
of the cluster will have settled down again. This will mean that the
luminosity boost at the core will make the core much brighter than the
surrounding temperature or luminosity should suggest.  This could make
the cluster look briefly like a cooling-flow cluster with a very high
accretion rate.  We find instantaneous mass accretion rates
$\dot{M}=L/(5kT/2\mu m_H)$ of up to 700\,$h^{-2}$\Msun\,yr$^{-1}$
similar to observed values in high redshift clusters (\citealt{EFA94};
\citealt{FaC95}; \citealt*{AFK96}; \citealt{AFE96}).

The current problem with cooling-flow clusters is that $t_{\rm{cool}}
\ll t_{\rm{H}}$ and so a lot of gas should have cooled, but this is
not observed \citep{KFT01,PPK01,PKP03}.  Unfortunately, even
considering the \citet{LMM99} result that cooling-flow clusters
usually occupy densely populated regions where clusters are more
likely to be undergoing mergers, this explanation cannot explain the
lack of cold gas in all cooling-flow clusters.  A {\it{ROSAT}} survey
\citep{PFE98} found that 70-90 per cent of clusters have cool cores.  By
their very nature the core mergers are short lived and so would be
rare.  However, it must be accepted that the accretion of subclumps
will continually feed the cores of clusters with low-entropy gas and
this must be a contributory factor to the resolution of the cool-core
problem.  The accretion of low-entropy material by clusters
will be investigated in a future paper.

\section*{Acknowledgements}

DRR is supported by a PPARC studentship.  STK is supported by
PPARC. These simulations were undertaken on the Cray T3E (RIP) at the
Edinburgh Parallel Computing Centre as part of the the Virgo
Consortium investigations of cosmological structure formation.
We would like to thank the referee whose comments helped to
considerably improve the content of the paper.

\bibliographystyle{mn2e}
\bibliography{clusters}

\label{lastpage}

\end{document}

%% file: clus.tab
\begin{table}
\caption{Cluster properties: identifying number; virial radius in
units of $h^{-1}$Mpc; virial mass in units of
$10^{14}h^{-1}$\Msun; virial temperature in units of keV/$k$;
sub-structure statistic; soft-band, emission-weighted
temperature excluding emission
within $50\,h^{-1}$kpc of the cluster centre in units of keV/$k$;
bolometric luminosity estimated from soft-band emission, excluding
emission from within $50\,h^{-1}$kpc of the cluster centre, in units
$10^{44}h^{-2}$erg\,s$^{-1}$; formation expansion factor as fit to the 
$M_{\rm vir}$ evolution--see Section~3.1. The clusters marked in bold
are ones which will be examined in detail below.}
\label{tab:clus}
\begin{center}
\begin{tabular}{rccccccc}
\hline
id& $R_{\rm vir}$& $M_{\rm vir}$& $T_{\rm vir}$& $T_X$& $L_{X}$& Sub.& $a_{\rm f}$\\
\hline
1 & 2.41& 17.85& 9.62& 7.15& 9.22& 0.11&  0.671\\
\bf2 & \bf1.80&  \bf7.49& \bf4.99& \bf4.68& \bf1.37& \bf0.05&  \bf0.529\\
3 & 1.79&  7.33& 4.63& 3.41& 1.13& 0.14&  0.718\\
4 & 1.78&  7.26& 5.09& 5.09& 1.73& 0.02&  0.688\\
5 & 1.67&  6.02& 3.75& 3.07& 0.43& 0.12&  0.890\\
6 & 1.62&  5.46& 4.04& 4.09& 1.04& 0.07&  0.557\\
7 & 1.62&  5.44& 3.61& 3.10& 0.38& 0.19&  0.756\\
\bf8 & \bf1.61&  \bf5.39& \bf4.27& \bf5.24& \bf2.63& \bf0.03&  \bf0.443\\
9 & 1.61&  5.34& 4.70& 3.94& 1.93& 0.15&  0.845\\
10& 1.60&  5.25& 4.09& 4.33& 1.36& 0.02&  0.459\\
\bf11& \bf1.60&  \bf5.22& \bf4.15& \bf4.19& \bf1.74& \bf0.08&  \bf0.731\\
12& 1.59&  5.13& 4.37& 4.58& 2.19& 0.03&  0.565\\
\bf13& \bf1.58&  \bf5.06& \bf5.38& \bf6.18& \bf3.12& \bf0.09&  \bf0.487\\
14& 1.56&  4.83& 3.83& 3.77& 0.52& 0.25&  0.774\\
15& 1.54&  4.71& 3.41& 4.11& 1.27& 0.06&  0.535\\
16& 1.54&  4.71& 3.15& 3.00& 1.14& 0.11&  0.777\\
17& 1.53&  4.61& 4.59& 4.19& 1.92& 0.06&  0.501\\
18& 1.53&  4.58& 3.62& 4.27& 0.84& 0.03&  0.699\\
19& 1.51&  4.43& 3.48& 4.38& 0.99& 0.05&  0.480\\
20& 1.35&  3.17& 3.13& 3.60& 1.02& 0.06&  0.401\\
\hline
\end{tabular}
\end{center}
\end{table}

%% file: majmerge.tab
\begin{table}
\caption{Number of clusters accreting various fractions of their 
material through major mergers. 
Also show is the mean angle (with respect to positive $\Tx$ evolution 
and no $\Lx$ evolution) in log space which the clusters move accross 
the $\Lx-\Tx$ plane (see Fig.~\ref{fig:ltevol}).}
\label{tab:majmerge}
\begin{center}
\begin{tabular}{lccccccc}
Fraction accreted & $\geq$0\% & $\geq$25\% & $\geq$50\% & $\geq$75\% \\
during mergers & $<$25\% & $<$50\% & $<$75\% & $\leq$100\%\\
\hline
Number of clusters & 4 & 4 & 8 & 4\\
\hline
Mean angle of \Lx-\Tx\ drift & -47$^\circ$ & 14$^\circ$ & 21$^\circ$ & 38$^\circ$\\
Mean dlog \Lx /dlog \Tx & -1.07 & 0.25 & 0.38 & 0.78\\
\end{tabular}
\end{center}
\end{table}